\documentclass[aps,prl,reprint,groupedaddress,showpacs]{revtex4-1}

\usepackage{graphicx}

\begin{document}


\title{Semiconductor ridge microcavity  source of quantum light at room temperature}


\author{X. Caillet$^1$, A. Orieux$^1$, A. Lema{\^i}tre$^2$, P. Filloux$^1$, I. Favero$^1$, G. Leo$^1$ and S. Ducci$^1$}
\affiliation{$^1$Laboratoire Mat{\'e}riaux et Ph{\'e}nom{\`e}nes Quantiques, CNRS-UMR 7162 Universit{\'e} Paris Diderot Case courrier 7021, 75205, Paris Cedex 13, France}
\affiliation{$^2$Laboratoire de Photonique et de Nanostructures, Route de Nozay, 91460 Marcoussis, France}


\date{\today}

\begin{abstract}
We experimentally demonstrate an integrated semiconductor ridge microcavity source of counterpropagating twin photons at room temperature in the telecom range. Based on parametric down conversion with a counterpropagating phase-matching, pump photons generate photon pairs with an efficiency of about $10^{-11}$ and a spectral linewidth of $0.3\: nm$ for a $1\: mm$ long sample. The indistiguishability of the photons of the pair are measured via a two-photon interference experiment showing a visibility of $85\:\%$. This work opens a route towards new guided-wave semiconductor quantum devices.
\end{abstract}

\pacs{42.50.Ex, 42.65.Lm, 42.65.Wi}

\maketitle


Entanglement is one of the most intriguing phenomena at the heart of quantum physics; in particular, entangled two-photon states have been used to confirm the foundations of quantum mechanics and constitute today one of the building blocks of quantum information and communication\cite{Tittel}. The most widely used way to produce entangled photon pairs is parametric down-conversion, in which one pump photon is annihilated into two lesser-energy correlated daughter photons which can be entangled in one or more of their degrees of freedom: frequency, polarization, momentum, and orbital angular momentum \cite{Multiparameter}. If we consider frequency, a narrow band pump beam, produces twin photons having perfectly anti-correlated frequencies. Recent developments in quantum information theory have arisen a growing interest on 'generalized' states of frequency correlation \cite{Grice} such as uncorrelated photons to guarantee their indistinguishability in linear optical quantum computation protocols \cite{LAFLAMME} or frequency correlated photons to improve clock synchronization\cite{Giovannetti}.  In this context, counterpropagating phase matching, in which a pump field impinges on top of a waveguide generating two counter-propagating wave-guided singal and idler beams, has been demonstrated to be the more flexible and versatile means to generate generalized states of frequency correlation \cite{Walton2}. 

On the other side, in the context of quantum information and communication a good deal of effort has been devoted to the miniaturization of quantum information technology on semiconductor chips, including micro-traps for ions\cite{Stick2006,Seidelin} and atoms\cite{Hansel2001}, and quantum-dot based sources of entangled photons\cite{Shields2007}. With respect to this last approach, parametric generation in semiconductor waveguides has the advantage of room-temperature operation and a highly directional emission, which dramatically enhances the collection efficiency. 

 \begin{figure}
 \includegraphics[width=8cm]{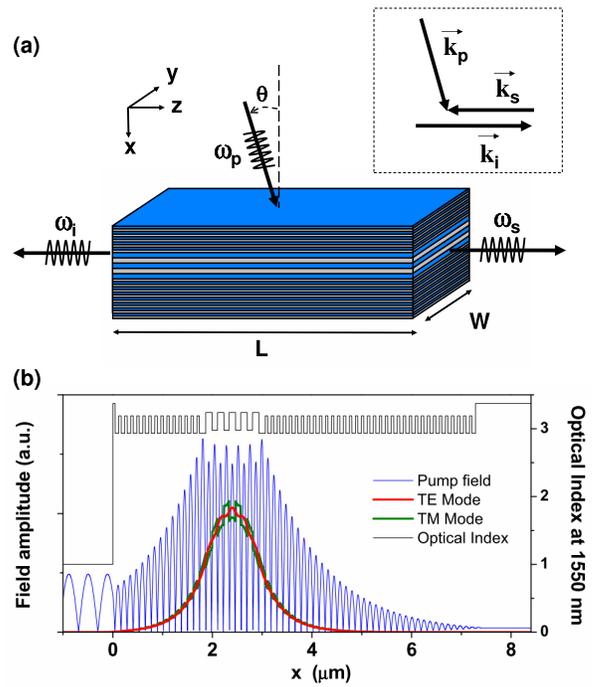}%
 \caption{Parametric generation of counterpropagating twin photons in a multilayer waveguide. a) Counterpropagating phase-matching scheme. The phase matching is automatically obtained in the z direction; Quasi Phase Matching (QPM) is provided by a periodic modulation of the nonlinear susceptibility in the waveguide core along the epitaxial direction. b) Amplitude of the interacting modes and optical index profile of  the microcavity. The pump field enhancement strongly improves the generation efficiency.
\label{fig1}}
 \end{figure}
 
In this letter we present a semiconductor ridge microcavity source of quantum light which combines the huge potential in terms of optoelectronic integration of semiconductor materials with the versatility in the production of the two photon state of the counterpropagating phase matching scheme.
A preliminary demonstration of a twin-photon source based on this principle has been reported in an $AlGaAs$ waveguide\cite{LancoPRL,Aspelmeyer2007}. In this experiment (Fig. \ref{fig1}a), a pump field impinges on top of the waveguide generating two counter-propagating, orthogonally polarised wave-guided twin photons through Spontaneous Parametric Down Conversion. The frequencies of the emitted fields are fixed by the energy  $\omega_p =\omega_s +\omega_i$) and momentum ($k_p \sin{\theta}  = n_sk_s - n_ik_i$) conservation, where $\omega_p$, $\omega_s$ and  $\omega_i$ ($k_p$, $k_s$ and $k_i$) are the frequencies (wave vectors) of pump, idler and signal;  $\theta$ is the angle of incidence of the pump beam, and $n_s$ and $n_i$ are the effective indices of the signal and idler modes. Momentum conservation in the epitaxial direction is satisfied by alternating $Al_{0.25}Ga_{0.75}As/ Al_{0.80}Ga_{0.20}As$ layers having different nonlinear susceptibilities to implement a Quasi Phase Matching (QPM) scheme. Since for each value of $\lambda$ and $\theta$  there is always a pair of photons satisfying these conservation laws, this geometry is also called auto-phase-matched\cite{Walton}. Moreover, as the three interacting beams travel in different directions, there is no need to filter the pump beam from the down-conversion and to separate the down-converted photons at a beam splitter, which represents an advantage with respect to collinear configurations. However, the performances reported in Ref. \cite{LancoPRL} were affected by low conversion efficiency and low signal/noise ratio due to photo-luminescence from the substrate, which prevented the utilization of the source for quantum optics experiments. A completely new design including two Distributed Bragg Reflectors (DBRs), one on the top and the other at the bottom of the waveguide, in order to create a microcavity for the pump beam allows to obtain a nearly standing wave inside the cavity. In this configuration, the internal amplitude of the pump field is much greater than outside; moreover, the presence of a lower DBR with a high reflection coefficient reduces the penetration of the pump field into the substrate, thus limiting the photo-luminescence noise\cite{Andronico}. Such vertical microcavity has been designed using the transfer matrix method. The vertical structure consists of 4.5 period $Al_{0.25}Ga_{0.75}As/Al_{0.80}Ga_{0.20}As$ QPM waveguide core, 41-period asymmetrical $Al_{0.35}Ga_{0.65}As/Al_{0.90}Ga_{0.10}As$ DBR (serving as lower cladding + bottom mirror) and 18-period asymmetrical $Al_{0.35}Ga_{0.65}As/Al_{0.90}Ga_{0.10}As$ DBR (serving as upper cladding + top mirror). The DBRs placed on both sides of the QPM region, play a double role: 1) waveguide cladding for the counterpropagating signal and idler; 2) mirror for the vertical cavity resonating at the pump wavelength. The sample was first grown by molecular beam epitaxy on a (100) $GaAs$ substrate, then chemically etched to create $2.5-3.5\: \mu m$ deep ridges with $5-6 \: \mu  m$ widths. The efficiency enhancement factor allowed by the vertical microcavity can be expressed as\cite{Andronico}:
\begin{equation}
\frac{\eta_{cavity}}{\eta_0}=\frac{2(1+n)^2}{\pi n}\frac{F}{1+\left|1+T_{down}/T_{up}\right|}
\end{equation}
where  $\eta_{cavity}$ ($\eta_{0}$) is the conversion efficiency defined as the ratio between the number of generated pairs and the number of pump photons, in the presence (absence) of the microcavity; $n$ is the mean effective index of the waveguide, $T_{up}$ ($T_{down}$) is the transmission coefficients of the upper (lower) mirror, and $F$ is the finesse of the cavity. In order to have an efficient process in the whole device, the resonance wavelength of the vertical microcavity must be the same on the entire length of the ridge; for this reason, in our design the upper value of $F$ is fixed by the quality of the homogeneity of the sample do to epitaxial growth. Fig. \ref{fig1}b shows the results of numerical simulations giving the amplitude profiles of the interacting fields at cavity resonance. 

\begin{figure}
 \includegraphics[width=8cm]{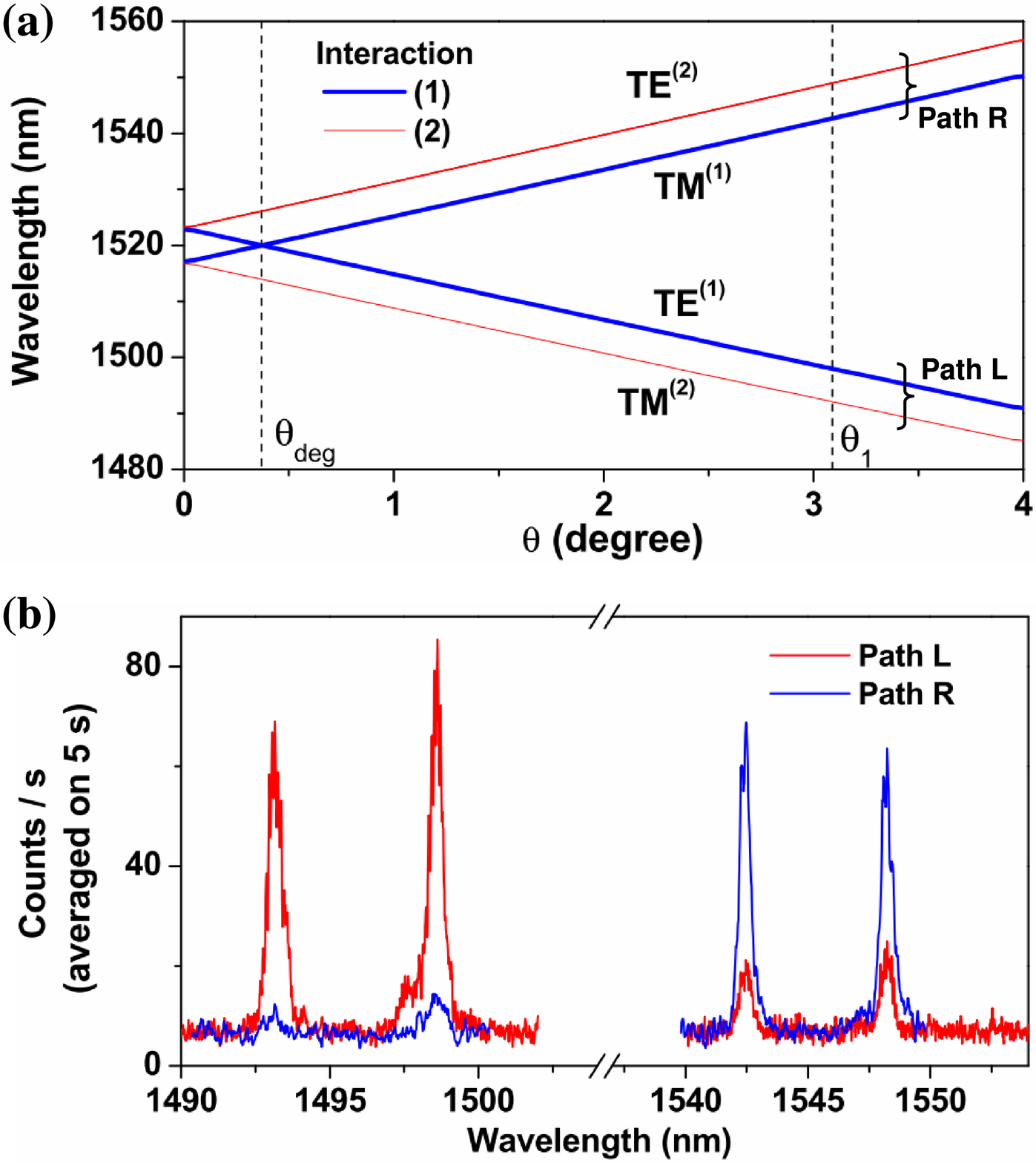}%
 \caption{Tunability curves and emission spectra.  a) Line: Calculated tuning curves as a function of the pump incident angle for a pump wavelength of $760\: nm$ and the structure described in the text. Dashed lines show the selected incident angles for the experiments.  b) Parametric fluorescence spectra for an angle of incidence of $\theta_1=3.1^{\circ}$ in the photon-counting regime. The peaks correspond to the four kinds of photons that can be generated via the two possible interactions. The background noise here is due to the dark counts of the detectors
\label{fig2}}
 \end{figure}

In our geometry, two equally probable interactions are possible: in the former (interaction 1), the photon copropagating with the $z$ component of the pump beam is TE polarised and the counterpropagating one; in the latter (interaction 2), the reverse occurs. Since the signal and idler central frequencies are determined by the conservation of energy and momentum in the $z$ direction, the incidence angle   of the pump beam provides a very convenient means to tune them. The X-shaped tunability curves of our source, typical of type II interactions, are shown in Fig. \ref{fig2}a. The degeneracy angle of interactions 1 and 2 are different, due to the modal birefringence of the waveguide. Note that a crucial feature of this device is the possibility of directly generating Bell states by simultaneously pumping at the two degeneracy angles. 

Fig. \ref{fig2}b shows the parametric fluorescence spectrum obtained with a TE polarised pump beam provided by a pulsed Ti:Sapphire laser, with wavelength  $\lambda_p=759.5\: nm$. The generated photons are collected from either facet of the waveguide with a microscope objective, spectrally analyzed with a monochromator, and then coupled into a fibered InGaAs single-photon avalanche photodiode. The spectra confirm the occurrence of the two processes predicted in Fig. \ref{fig2}a and demonstrate the possibility of direct generation of polarisation-entangled states. The amplitude difference in the observed signal is due to the fact that the long wavelength photons are collected after their reflection on the opposite facet. An antireflection coating on both facets of the sample would allow an automatic separation of the photons of each pair and their direct coupling into two optical fibres, through standard pigtailing. In this experiment, the bandwidth of the generated photons stems from the convolution of the pump spectrum ($0.3\: nm$ bandwidth), the phase matching band ($0.3\: nm$ bandwidth for a sample of length $L=1\:mm$) and of the monochromator resolution ($0.1\: nm$). In the case of a monochromatic pump beam, the spectrum of the down-converted photons is given by the usual function $sinc^2 (\Delta k L/2)$. One of the main advantages of the counterpropagating geometry arises from the rapid increase of  $\Delta k$ when one moves away from perfect phase matching, which leads to a bandwidth of the downconverted photons that is one to two orders of magnitude narrower than for collinear geometries: this lends itself to long-distance propagation in optical fibres, with a negligible chromatic dispersion. Measuring the amount of detected photons, we estimate the brightness of our twin photon source to be around $10^{-11}$, representing an enhancement of at least two orders of magnitude with respect to the device presented in ref. \cite{LancoPRL}. Moreover, the consequent improvement of the signal/noise ratio obtained thanks to the integration of the vertical microcavity makes the source suitable for quantum optics applications.  

\begin{figure}
 \includegraphics[width=8cm]{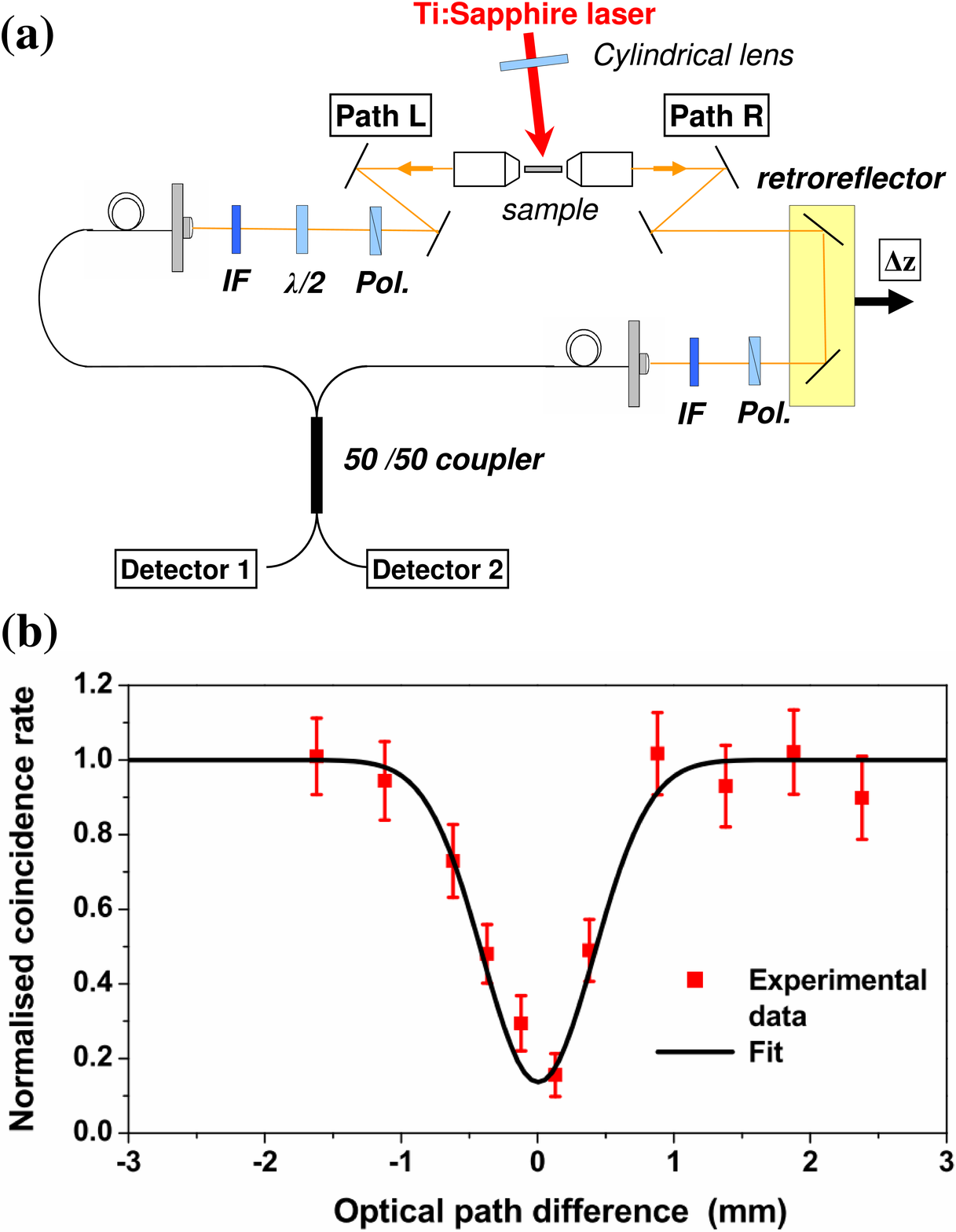}%
 \caption{Two-photon interference set-up and Hong-Ou-Mandel dip.  a) Sketch of the experimental setup used for the two photon interference. The two photons generated in the selected interaction are combined in a fibered 50/50 coupler, where quantum interference occurs. The polarisation of one of the two photons is rotated of $90^{\circ}$ with a half-wave plate. A retroreflector placed in one arm of the interferometer is used to adjust the relative delay of the two photons. b) Net coincidence counting rate (total counts - accidental counts) as a function of the relative length of the  two arms. The accidental coincidences accounts for $14\:\%$ of the total coincidences outside the dip. The error bars are determined by the Poisson distribution (the square-root sum of the total coincidence counts) and the solid line is the numerical fit.
\label{fig3}}
 \end{figure}

Since polarisation-entangled photon pairs require the indistinguishability of the two photons for any degree of freedom but polarisation, a HOM experiment can be used to test the quality of the quantum properties of our source. In this type of experiment, two indistinguishable photons enter a 50/50 beam splitter at the same time; the destructive interference makes them exit the device through the same output, thus inducing a dip in the coincidence histogram \cite{MandelHong}.  
The visibility of the dip gives informations on the quality of indistinguishability and thus on the potential amount of entanglement, while the width corresponds to the coherence length of the interfering photons. The application of this test to our source requires that the two photons of the pair are turned indistinguishable by rotating the polarization of one of the two photons of $90^{\circ}$.
Fig. \ref{fig3}a shows a sketch of the experimental setup for the two-photon interference. The waveguide is pumped with a pulsed Ti:Sapphire laser ($150\: ns$ pulses, $3\: kHz$ repetition rate, $10\: W$ peak power). The pump wavelength is $760\: nm$ and its linewidth $0.3\: nm$. The beam is collimated with a cylindrical lens on top of the waveguide ridge, on a length of $0.65\: mm$, at $\theta_1=0.37^{\circ}$, the degeneracy angle of interaction 1. This leads to the generation of $10$ pairs per pump pulse, equally distributed on the two kinds of interactions; the pairs generated via interaction 1 are selected with two polarisers. A retroreflector placed in one arm of the interferometer is used to adjust the relative delay between the two photons and a half-wave plate is used to make their polarisations parallel.  The photons recombine onto a fibered 50/50 beam splitter and the signal emerging from its two outputs is detected. The overall transmission coefficient of the interferometer is $12.5\:\%$ ($70\:\%$ waveguide facets, $70\:\%$ microscope objectives, $50\:\%$ interference filters, $50\:\%$ fibre beam splitter). The interference filters, centred at $1520\: nm$ with a bandwidth of $10\: nm$, are used to reduce the luminescence noise (which is a white noise of the order of 0.05 photons per nanometre per pump pulse, over a width of a few hundred nanometres). The photons are detected via two InGaAs single-photon avalanche photodiodes (Id 201 from ID Quantique); in our operating conditions their detection efficiency is $20\:\%$ and the dark count rate is $20$ counts/s. This set-up typically allows to detect a signal of $400$ counts/s, with a luminescence noise of about $20$ counts/s.

Fig. \ref{fig3}b reports the dip in the coincidence counts observed as a function of the optical path length difference between the two arms, once the accidental counts have been removed: this dip is a clear signature of the destructive interference between the two photons. The solid line shows the fit between our data and the theoretical expression of the HOM dip \cite{Wang}:
\begin{equation}
N_c=1-V\exp{\left(-\frac{\pi^2}{\log{2}}\left[\frac{\delta z \Delta\lambda}{\lambda^2}\right]^2\right)}
\end{equation}
 where $N_c$ is the normalised coincidence rate, $V$ the visibility,  $\delta z$ is the optical path difference,  $\lambda$ is the degeneracy wavelength, and   $\Delta\lambda$ is the full width at half maximum spectral intensity. The two fitting parameters are  $\Delta\lambda$  and $V$; for the first one we obtain $0,53\: nm$, in excellent agreement with our numerical simulations, while for the second one we obtain $85\:\%$, which is the same as what has been recently obtained in periodically poled $LiNbO_3$ waveguides\cite{Martin}. We note that this value is obtained without the use of filters to reduce the spectral bandwidth of the photons emitted by the source. The main cause of the imperfect visibility in our experiment is that the waveguide facets are not anti-reflection coated and have a reflectivity $R=30\:\%$. Note that, in our set up, a coincidence event can be not only due to two photons directly transmitted by the facets, but also by one photon directly transmitted and one photon having experienced two reflections before leaving the waveguide. In the latter case, the path difference for the two photons is not the same as for the first case, so these photons do not contribute to the dip. If we consider the first case to have a probability $1$, the second case has a probability $2R^2$, since the twice-reflected photon can be first reflected at either the left or the right facet. The visibility is then given by  $V=1/(1+2R^2)=85\pm3\:\%$, which is in perfect agreement with our experimental results.

These results constitute the first demonstration of two-photon interference obtained with a semiconductor source at room temperature. They pave the way to the demonstration of a few interesting features associated to the counterpropagating geometry, as the direct generation of polarisation entangled Bell states, or the control of the generated two-photon state via an appropriate choice of the spatial and spectral pump beam profile\cite{Grice,Walton2,Caillet} . The efficiency of this room temperature working device, along with the high-quality quantum properties of the generated photons and their telecom wavelength, makes this source a serious candidate for integrated quantum photonics.

\begin{acknowledgments}
This work has been partially funded by the Germaine de Stael programme of Parternariat Hubert-Curien.
\end{acknowledgments}


%

\end{document}